\documentclass[acmsmall,screen,nonacm,review=false]{acmart}

\setcopyright{none}
\authorsaddresses{%
  \textsuperscript{*}Alex Flesher and Nitheesh Kodarapu contributed equally.\\
  Contact: Stephen Abkin, \href{mailto:stephen122204@tamu.edu}{stephen122204@tamu.edu}.\\ Code and experiment records: \url{https://github.com/stephen122204/mantra-energy/tree/paper-1}}

\newcommand{\sourcepath}{.}
\graphicspath{{./}}

\usepackage{booktabs}
\usepackage{tabularx}
\usepackage{multirow}
\usepackage{array}
\usepackage{tikz}
\usetikzlibrary{arrows.meta,positioning,calc,decorations.pathreplacing,fit}
\usepackage[capitalise,noabbrev]{cleveref}
\crefname{appendix}{Appendix}{Appendices}
\Crefname{appendix}{Appendix}{Appendices}
\usepackage{etoolbox}
\usepackage{subcaption}
\usepackage{flafter}                   %
\usepackage{placeins}                  %
\usepackage{colortbl}
\definecolor{tablehead}{HTML}{EFEFEF}
\definecolor{favorfour}{HTML}{F8E3CF}   %
\definecolor{favoreight}{HTML}{DCE7F4}  %
\newcommand{\headrow}{\rowcolor{tablehead}}
\newcommand{\tablegroup}[2]{\multicolumn{#1}{@{}l}{\textit{#2}}\\}
\AtBeginEnvironment{tabular}{\small}   %
\AtBeginEnvironment{tabularx}{\small}
\definecolor{insightbg}{HTML}{ECECEC}
\newcommand{\insight}[2][Insight]{\par\medskip\noindent\colorbox{insightbg}{\parbox{\dimexpr\linewidth-2\fboxsep\relax}{\textbf{#1.}\ #2}}\par\medskip}

\newcommand{\Etok}{E_{\mathrm{tok}}}   %
\newcommand{\Einf}{E_{\mathrm{inf}}}   %
\newcommand{\Eop}{E_{\mathrm{op}}}     %
\newcommand{\Qfour}{\textsc{Q4\_K\_M}}
\newcommand{\Qeight}{\textsc{Q8\_0}}
\newcommand{\BF}{\textsc{BF16}}

\newcommand{\rhotok}{\rho_{\mathrm{tok}}}

\newcommand{\rhoop}{\rho_{\mathrm{op}}}

\makeatletter
\patchcmd{\@mktitle@i}{\raggedright}{\centering}{}{\PackageError{arxiv}{Title centering failed}{}}
\makeatother

\makeatletter
\renewcommand{\@mkauthors@i}{%
  \global\setbox\mktitle@bx=\vbox{%
    \unvbox\mktitle@bx\par\medskip
    \begingroup\centering\normalfont\normalsize
    Stephen Abkin, Aldric Yang, Alex Flesher\textsuperscript{*}, Nitheesh Kodarapu\textsuperscript{*}\\[2pt]
    Deepak Panigrahy, Aakash Tyagi, David Kebo Houngninou\\[5pt]
    Texas A\&M University\par
    \endgroup\medskip}}
\makeatother

\newcommand{\preprintpagestyle}{%
  \fancyhf{}
  \fancyfoot[L]{\footnotesize Preprint}
  \fancyfoot[C]{\footnotesize\thepage}
  \renewcommand{\headrulewidth}{0pt}
  \renewcommand{\footrulewidth}{0pt}}
\AtBeginDocument{%
  \fancypagestyle{standardpagestyle}{\preprintpagestyle}
  \fancypagestyle{firstpagestyle}{\preprintpagestyle}
  \pagestyle{standardpagestyle}}

\begin{document}

\title[Task Completion Energy with Quantized Language-Model Agents]{Measuring the Energy Cost of Completing Tasks with Quantized Language-Model Agents}

\author{Stephen Abkin}
\author{Aldric Yang}
\author{Alex Flesher}
\author{Nitheesh Kodarapu}
\author{Deepak Panigrahy}
\author{Aakash Tyagi}
\author{David Kebo Houngninou}
\affiliation{\institution{Texas A\&M University}\city{College Station}\state{Texas}\country{USA}}
\renewcommand{\shortauthors}{Abkin et al.}

\begin{abstract}
Quantization reduces the memory and energy needed to serve large language models (LLMs) and AI agents for coding and data analysis can now run on locally hosted LLMs. In practice, this benefit is commonly benchmarked per output token. Yet an agent calls its model repeatedly, runs tools, and revises its work, and quantization can change both how much the agent generates and whether it succeeds. Existing energy studies either compare quantized models on single requests or measure agents at one fixed precision, overlooking how a change in precision plays out across a complete agent task.

To address this gap, this paper presents the first characterization of the task-level energy trade-offs of model quantization in locally hosted agents. We develop 29 agent tasks with automatic pass/fail checks in six families, ranging from single-answer SQL analytics to multi-step repository bug fixing, and measure CPU and GPU energy per output token and per completed task for 4-bit and 8-bit versions of two open-weight models on bare-metal NVIDIA A100 nodes with llama.cpp.

Our findings reveal that 4-bit precision generally saves energy per token, though whether it saves energy per completed task is task dependent and changes with the model and the execution budget (i.e., limits on turns, tokens, and time). We trace this discrepancy to the extra tokens needed to complete a task in full, which translates into using up to 3.5 times as much energy per completed task. These findings offer actionable guidance for developers and operators choosing model precision for agents, showing how tokens per completed task, set against the per-token saving, reveal whether lower precision completes the work with less energy.
\end{abstract}

\renewcommand{\keywordsname}{\textbf{\textit{Keywords}}}

\maketitle

\section{Introduction}
\label{sec:intro}

Quantization is becoming a practical way to run LLMs locally on a single GPU instead of a provider's cluster. Models such as Qwen3~\cite{yang2025qwen3} and Devstral~\cite{devstralmodelcard} can now drive coding and data agents on one GPU, served by engines such as llama.cpp~\cite{llamacpp} from files that community repositories publish at many precisions, and methods such as Samsung's LittleBit already compress a 13-billion-parameter model into under 1\,GB~\cite{lee2025littlebit}. Capable agents on consumer hardware are still emerging, but agents on a single GPU that an organization owns are already a reality. There, the energy of every completed task arrives on the electricity bill. An operator then chooses the precision of the model's weights to fit its memory and energy budget, and quantization appears to make this choice easy. A 4-bit file needs less memory than an 8-bit one, usually generates each token with less energy, and often keeps accuracy close to the unquantized model~\cite{jin2024comprehensive,giagnorio2026quantization}. Benchmarks such as TokenPowerBench report exactly this energy per output token~\cite{niu2026tokenpowerbench,samsi2023words,wilhelm2025energypertoken}. Yet the useful output of an agent is a completed task, not a token. Completing a task can take many model calls, tool runs, and revisions, and a quantized model may need more of them or fail more often. This raises the question, is the cheaper precision per token also the cheaper one per completed task?

An agent turns a per-token saving into task energy only through its behavior. Each extra turn (i.e., model response) adds a model call over a longer context, a response with a malformed tool call wastes a turn, and a wrong submission discards the whole attempt. Quantization can add to these costs. \citet{jang2026flatscore} show that quantized agents make more failed tool calls and need more model responses and tool calls before finishing, and \citet{pham2026agentstop} find that failed attempts carry a substantial energy cost of their own. Because these costs accumulate over the whole task, the task itself is the natural unit of cost. In guidance published in September 2026, Anthropic~\cite{osmani2026taskcost} makes this argument in terms of monetary cost, recommending that developers of coding agents measure what a task costs rather than its price per token, since models with equal per-token prices can cost very different amounts on the same task. Energy behaves the same way, even when both precisions always succeed. 

\paragraph{\upshape\bfseries Motivation} Our motivation stems from quantization being one of the most accessible mechanisms a developer has for running agents on their own hardware. Choosing a 4-bit instead of an 8-bit quantized model needs no new hardware and no change to the agent, and it lowers both memory and per-token energy. An agent runs the same kinds of tasks many times, so the energy cost of one precision choice repeats with every task it completes. If the per-token saving carries through to completed tasks, quantization is close to a free energy saving for local agents. If agent behavior erases it, a developer could end up deploying the costlier precision while per-token benchmarks point the other way. This calls for measuring quantization on the tasks an agent completes, where its energy is actually spent.

\paragraph{\upshape\bfseries Goal} To our knowledge, a task-level understanding of quantization's energy is lacking. Prior work reports energy per token or per request~\cite{husom2025sustainable,shi2025qmeter,chung2025mlenergy,chung2026joules}, measures agents at a single precision~\cite{ifath2026multirequest,luo2026blackwell,pham2026agentstop}, or compares precisions by correct answers on single-turn queries or with energy calculated rather than read from hardware counters~\cite{saadfalcon2025ipw,gao2026tokenarena}. To address this gap, we conduct a controlled empirical study that measures each precision's energy both per output token and per completed task, and asks whether the two measures favor the same precision. We report the results as four findings for developers and operators choosing model precision for agents, and address four research questions, which \cref{sec:results} answers in order.

\begin{description}
\item[RQ1:] Do energy per token and energy per completed task favor the same precision?
\item[RQ2:] Where they differ, how much extra energy per completed task does choosing by energy per token cost?
\item[RQ3:] What in the agent's behavior accounts for the extra energy, such as how much it generates or how often it succeeds?
\item[RQ4:] How do the execution budget, the model, and the task set change which precision is cheaper per completed task, and is an unquantized 16-bit model (which needs twice the GPU memory) ever cheaper than the quantized files?
\end{description}

We run the 4-bit \Qfour{} and 8-bit \Qeight{} files\footnote{\Qfour{} and \Qeight{} are llama.cpp quantization formats that store most weights in 4 and 8 bits (\cref{sec:background:quant}). We call them 4-bit and 8-bit for short.} of Qwen3-30B-A3B~\cite{qwen3modelcard} on 29 verifiable tasks (i.e., an automatic check decides whether each attempt succeeds) and compare the precision that each measure favors to address RQ1. Where the two measures favor different precisions, which we call a \emph{disagreement,} we measure the extra energy per completed task to answer RQ2. We trace each difference to how much the agent generates, how often it succeeds, and how much energy it spends outside model calls to answer RQ3. To address RQ4, we repeat the comparison with an expanded budget, a 16-bit reference, Devstral Small~2~\cite{devstralmodelcard}, and held-out tasks from the BIRD text-to-SQL and EvalPerf code-efficiency benchmarks~\cite{li2023bird,liu2024evalperf}.

\paragraph{\upshape\bfseries Summary of findings} 4-bit uses 6--16\% less energy per output token than 8-bit on average over the 29 tasks for both models, so per-token benchmarks favor it almost everywhere (\cref{sec:results:rq1}). Yet on some everyday coding and data-analysis tasks, 4-bit uses more energy per completed task, beyond run-to-run variation. On such tasks, 4-bit uses a median of 66\% more energy per completed task than 8-bit, and in the worst supported case, 4-bit needs 3.5 times as much energy as 8-bit to complete the same task (\cref{sec:results:rq2}). Where the two measures disagree, the extra energy comes from extra tokens, as the precision that is cheaper per token needs a median of 1.6--2.0 times as many tokens per completed task. These tokens arise from malformed tool calls and unnecessary reruns of tests that had already passed, which prolong attempts, as well as wrong answers that cause attempts to fail (\cref{sec:results:rq3}). As tasks scale in difficulty, the precision choice becomes more concerned with completion rate, and which precision completes more depends on how much context and how many turns the agent is given (\cref{sec:results:rq1,sec:results:rq4}). These findings hold through a second model, whose costly tasks differ from the first, and the patterns of our task families carry over to held-out benchmark problems, correctly predicting where 4-bit is cheaper per completed task. An unquantized 16-bit model is never clearly cheaper than the quantized files (\cref{sec:results:rq4}). We propose that precision should be chosen by measuring energy per completed task for the model and tasks an agent will run, under the circumstances in which it will be deployed.

\paragraph{\upshape\bfseries Contributions} We summarize our main contributions as follows:
\begin{itemize}
\item We characterize the task-level energy of quantized models in locally hosted agents across 29 verifiable tasks in six families and held-out benchmarks (\cref{sec:method}).
\item We build a measurement harness that runs the agent, attributes CPU and GPU energy to each model and tool call, checks each attempt with hidden verifiers, and charges failed attempts to completions (\cref{sec:metrics,sec:method}).
\item We empirically show that agent behavior can outweigh the per-token saving, making 4-bit costlier per completed task on some tasks, by a margin beyond run-to-run variation (\cref{sec:results}).
\item We derive an identity that splits energy per completed task into energy per token and tokens per completed task, so a lower precision saves energy roughly as long as its extra tokens per completed task stay below its per-token saving (\cref{sec:metrics:identity,sec:discussion:implications}).
\end{itemize}

\FloatBarrier
\section{Background and Related Work}
\label{sec:background}

\subsection{Quantized models for local inference}
\label{sec:background:quant}

\begin{figure}[t]
\centering
\includegraphics[width=\linewidth]{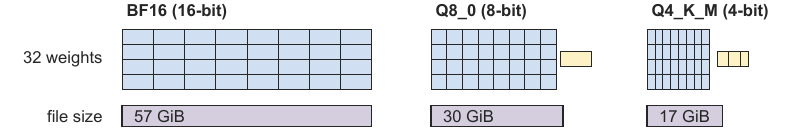}
\caption{Each blue cell is one weight, drawn with a width proportional to its stored bits, and each column holds the same 32 weights of Qwen3-30B-A3B. Yellow cells are the values that each quantized block shares to rescale its weights (one 16-bit scale in \Qeight{}, and a scale, an offset, and a share of a larger group's scale and offset in \Qfour{}), so they add storage beyond the bit width. The bars give the measured file sizes on one scale, which decide whether a model fits on one GPU. File sizes are in GiB ($2^{30}$ bytes), and GPU capacities are in decimal GB.}
\Description{Three columns labeled BF16 16-bit, Q8\_0 8-bit, and Q4\_K\_M 4-bit. Each shows the same grid of 32 weights, with cells half as wide at 8 bits and a quarter as wide at 4 bits, plus small yellow cells for shared scales next to the quantized grids. Bars below give file sizes of 57, 30, and 17 GiB on one scale.}
\label{fig:precision}
\end{figure}

Local deployments commonly serve models with llama.cpp~\cite{llamacpp} from GGUF files\footnote{GGUF is the single-file model format of llama.cpp. One file stores the weights together with the tokenizer and the model metadata, so it fully specifies what the engine loads.}~\cite{gguf}, and community repositories publish each model at many precisions. A precision label such as \Qfour{} names a quantization scheme, not only a bit width. \Qfour{} stores most tensors at 4 bits in blocks that share a scale and an offset, and it keeps some attention and feed-forward tensors at 6 bits~\cite{kawrakow2023kquants}. Our 4-bit files were also calibrated on sample text, so weights whose inputs are largest on that text are rounded most carefully~\cite{bartowski2025qwen3gguf,bartowski2025devstralgguf}. \Qeight{} rounds weights to 8 bits in blocks of 32 with one scale per block and needs no calibration~\cite{llamacppquantize}. A \BF{} file holds the original, unquantized weights in 16-bit bfloat16. Because of the shared scales and the 6-bit tensors, each quantized file stores more bits per weight than its label suggests (8.5 for \Qeight{} and 4.88 for our 4-bit Qwen file). The 4-bit Qwen file is thus about 30\% the size of the 16-bit one rather than a quarter (\cref{fig:precision}). The Devstral files use the same quantized layouts. Devstral is released in 8-bit floating point, so it has no original 16-bit weights.

Quantized models, including 4-bit ones, often stay close to their unquantized versions on general and code-generation benchmarks~\cite{jin2024comprehensive,li2024evaluating,giagnorio2026quantization,wei2023greener}. The loss is larger on long-context inputs~\cite{mekala2025longcontext} and on realistic agent applications such as simulated science experiments and text games, where 4-bit quantization lowered accuracy by 10 to 15\%~\cite{dong2025acbench}. Whether quantization lengthens outputs depends on its severity and on the model~\cite{liu2025quantizationreasoning}. The speed and energy saved by a lower precision also depend on the hardware, kernels, and batching, including whether dequantization is fused into the matrix multiplication~\cite{benazir2025applesilicon,chung2026joules,delavande2026understanding,rajput2024benchmarking}.

\subsection{From energy per token to energy per completed task}
\label{sec:background:units}
\label{sec:background:agents}
\label{sec:background:position}

\paragraph{\upshape\bfseries Units beyond the token} Energy per output token is the common unit of inference benchmarks~\cite{niu2026tokenpowerbench,samsi2023words,wilhelm2025energypertoken}, and studies that compare quantization methods report it beside accuracy without combining the two~\cite{husom2025sustainable,shi2025qmeter}. Several studies document its limits for whole requests. ML.ENERGY reports energy per response because models differ in verbosity~\cite{chung2025mlenergy,chung2026joules}. Energy per request grows with response length~\cite{husom2024price,wilkins2024offline}, and \citet{luo2026blackwell} found that enabling a model's thinking mode generated 21 to 75\% more tokens per problem at nearly unchanged energy per token. The measured boundary (i.e., which hardware components and periods the energy covers) matters as well. GPU-only telemetry missed 41 to 45\% of system energy on one agent server~\cite{luo2026blackwell}.

\paragraph{\upshape\bfseries Charging energy to completions} Several measures already charge cost to successful outcomes. The cost-of-pass of \citet{erol2025costofpass} divides the monetary cost of a model's attempts by its success rate. The energy per successful goal of \citet{panigrahy2026epg} divides the energy of all attempts, including failed ones, by the number of successes, and our energy per completed task (\cref{sec:metrics}) applies this goal-level accounting to the CPU packages and GPU board of the machine that runs the model and its tools. Intelligence per Watt relates accuracy on single-turn queries to measured energy across 16-, 8-, and 4-bit models~\cite{saadfalcon2025ipw}, and TokenArena ranks endpoints by estimated joules per correct answer~\cite{gao2026tokenarena}. We build on these measures by computing energy per output token and energy per completed task on the same agent attempts, which shows when the two favor different precisions of one model.

\begin{figure}[t]
\centering
\begin{subfigure}[b]{0.74\linewidth}\centering
\includegraphics{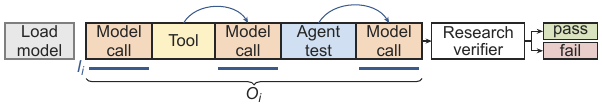}
\caption{One attempt and its measured periods.}
\end{subfigure}\hfill
\begin{subfigure}[b]{0.245\linewidth}\centering
\includegraphics{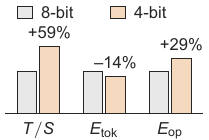}
\caption{Business query.}
\end{subfigure}
\caption{Energy accumulates over the model calls, tools, and waiting of the task window, and in the business query of our study (i.e., an SQL query that summarizes sales by customer segment), 4-bit is cheaper per token but costlier per completed task. (a) An example attempt, not to scale. Blue marks are model calls, whose energy is $I_i$, and the brace spans the task window, whose energy is $O_i$. Model loading and the research verifier (i.e., the study's final check of the result) are measured separately and excluded from $O_i$. (b) Ten verified completions per precision of the business query at one million rows, showing tokens per completed task ($T/S$), energy per output token ($\Etok$), and energy per completed task ($\Eop$), with bar heights normalized to the 8-bit value and labels giving the 4-bit change.}
\Description{Panel a shows an attempt as a row of blocks after a model-load block, alternating model calls with a tool and an agent test, whose results return by curved arrows to the next model call. Blue marks under the model calls are labeled I sub i, and a brace under the whole row is labeled O sub i. The last model call leads to a research verifier with pass and fail outcomes. Panel b shows paired vertical bars for 8-bit and 4-bit, labeled T over S, E tok, and E op. 4-bit has 59 percent more tokens per completion, 14 percent less energy per token, and 29 percent more energy per completion.}
\label{fig:mechanism}
\end{figure}

\paragraph{\upshape\bfseries Agents and precision} For agents, \citet{ifath2026multirequest} characterize multi-request workflows under different serving setups, and \citet{luo2026blackwell} attribute a coding agent's energy to components and phases, both at a fixed precision. The two closest studies each hold one side of our question fixed. \citet{pham2026agentstop} quantify the energy of failed runs of a local 4-bit agent, and \citet{jang2026flatscore} vary precision without measuring energy, finding that a quantized agent can succeed as often while making more failed tool calls. We combine the two sides, varying precision and measuring energy on the same attempts, to test whether the extra tokens and failed attempts that a lower precision induces outweigh its per-token saving.

\section{Three Energy Measures}
\label{sec:metrics}

To see when energy per token and energy per completed task favor different precisions, we compute both on the same agent attempts, together with a third measure between them, model-call energy per completed task (\cref{sec:metrics:defs}). An algebraic identity (\cref{eq:product}) links the three and shows what can change the preferred precision.

\subsection{Attempts and the three measures}
\label{sec:metrics:defs}

An \emph{attempt} is one run of the agent on one task at one precision (i.e., one model file served entirely from the GPU). The attempt starts from a newly loaded model and ends when the agent submits a result or reaches a limit of its execution budget (\cref{sec:method:budgets}). During the attempt, the agent alternates between model calls and tool calls and uses each result to decide its next action. After the attempt, a deterministic research verifier checks the submitted result. It is the study's own final check, hidden from the agent and separate from the tests that the agent runs. An attempt that the research verifier accepts is a \emph{verified completion}, and every other attempt is a failure.

\Cref{fig:mechanism}a shows how an attempt accumulates energy, which we measure on the CPU packages (i.e., processor sockets) and GPU board (i.e., the whole graphics card) of the machine that runs the model and its tools. The \emph{model-call (inference) energy} $I_i$ of attempt $i$ is the energy recorded during its model calls, in which the model reads its input tokens (the prompt) and generates $T_i$ output tokens. The \emph{operational energy} $O_i$ is the energy of the whole task window, from the loaded model to the end of the task. It adds every tool call, including agent-requested tests, and the time the machine waits between calls. Model loading and the research verifier are measured separately and excluded from $O_i$ (\cref{sec:method:energy}). We exclude loading because a deployed server loads the model once for many tasks, and \cref{sec:results:rq1} reports the worst case, one load per attempt. We write $s_i=1$ if attempt $i$ is a verified completion and $s_i=0$ otherwise.

For one precision on one task, we sum four quantities over its $n$ valid attempts, those with a complete and undisturbed energy measurement (\cref{sec:method:validity}). They are the model-call energy $I=\sum_i I_i$, the operational energy $O=\sum_i O_i$, the output tokens $T=\sum_i T_i$, and the number of verified completions $S=\sum_i s_i$. Each measure divides one total by another,
\begin{equation}
\label{eq:defs}
\Etok = \frac{I}{T}, \qquad
\Einf = \frac{I}{S}, \qquad
\Eop = \frac{O}{S}.
\end{equation}
Lower values are better for all three. Because $\Einf$ and $\Eop$ divide by verified completions, they also reflect how reliably a precision succeeds, which $\Etok$ does not reflect.
\begin{itemize}
\item \textbf{Energy per token, $\Etok$.} The model-call energy per output token, the unit that per-token benchmarks report~\cite{niu2026tokenpowerbench,samsi2023words}.
\item \textbf{Model-call energy per completed task, $\Einf$.} It shares its numerator with $\Etok$ and its denominator with $\Eop$, so it shows which of the two steps moves the preferred precision (\cref{sec:metrics:selection}).
\item \textbf{Energy per completed task, $\Eop$.} The operational energy per verified completion. It charges every attempt, failed or not, to the completions, following the energy per successful goal of \citet{panigrahy2026epg}.
\end{itemize}
All three are ratios of sums, so each attempt counts in proportion to its energy and tokens, and one long attempt can outweigh several short ones. A precision that never completes a task ($S=0$) has no $\Einf$ or $\Eop$ on that task, and \cref{sec:metrics:selection} covers such tasks.

\subsection{The token factor, the boundary factor, and the break-even rule}
\label{sec:metrics:identity}

Energy per completed task is energy per token scaled by the tokens each completion takes and by the energy outside the model calls. Multiplying and dividing $\Einf$ by $T$ gives $\Einf = \Etok \cdot T/S$, where $T/S$ is the number of output tokens per verified completion, counted over all attempts. Multiplying and dividing $\Eop$ by $I$ gives $\Eop = \Einf \cdot O/I$, where $O/I \ge 1$ says how far the energy of the whole task window exceeds that of its model calls. To compare the 4-bit and 8-bit precisions of one model, we mark each total by a subscript 4 or 8 and divide each 4-bit measure by its 8-bit value. We call $\rhotok = E_{\mathrm{tok},4}/E_{\mathrm{tok},8}$ the \emph{per-token ratio} and $\rhoop = E_{\mathrm{op},4}/E_{\mathrm{op},8}$ the \emph{per-task ratio}, and a ratio below one favors 4-bit. Combining the two steps and dividing the 4-bit value by the 8-bit one gives
\begin{equation}
\label{eq:product}
\rhoop \;=\; \rhotok \,\times\,
\underbrace{\frac{T_4/S_4}{T_8/S_8}}_{\text{token factor}}
\,\times\,
\underbrace{\frac{O_4/I_4}{O_8/I_8}}_{\text{boundary factor}} .
\end{equation}
The \emph{token factor} compares tokens per completed task. The \emph{boundary factor} compares how much each precision's energy grows when the measured boundary widens from the model calls to the whole task window. $\Etok$ and $\Eop$ favor different precisions exactly when $\rhotok$ and $\rhoop$ lie on opposite sides of one.

Tokens per completed task has two parts, the tokens of an average attempt and how many attempts each completion takes. With the completion rate $p=S/n$ and the mean output tokens per attempt $\bar{T}=T/n$, we have $T/S=\bar{T}/p$, and the token factor splits into
\begin{equation}
\label{eq:tokens}
\frac{T_4/S_4}{T_8/S_8} \;=\; \underbrace{\frac{\bar{T}_4}{\bar{T}_8}}_{\text{length factor}}\,\times\,\underbrace{\frac{p_8}{p_4}}_{\text{completion factor}}.
\end{equation}
The \emph{length factor} grows when 4-bit attempts are longer, and the \emph{completion factor} $p_8/p_4$ grows when 4-bit completes less often.

Splitting $O$ by outcome shows how failed attempts enter $\Eop$. Let $\bar{O}^{+}$ and $\bar{O}^{-}$ be the mean operational energy of the $S$ successful and the $n-S$ failed attempts, respectively. For $0<S<n$,
\begin{equation}
\label{eq:failures}
O = S\,\bar{O}^{+} + (n-S)\,\bar{O}^{-},
\qquad\text{so}\qquad
\Eop = \frac{O}{S} = \bar{O}^{+} + \bar{O}^{-}\,\frac{n-S}{S} = \bar{O}^{+} + \bar{O}^{-}\,\frac{1-p}{p}.
\end{equation}
Each completion carries the energy of one successful attempt plus a share of the failed ones, and the weight $(1-p)/p$ grows quickly as the completion rate falls. At $p=1/2$ each completion also pays for one failed attempt, and at $p=1/4$ it pays for three.\footnote{When attempts are independent and identically distributed, $O/S$ also estimates the expected operational energy spent until the first success~\cite{panigrahy2026epg,wald1944cumulative}.}

Even the per-token ratio depends on what the agent did. Each precision's $\Etok$ is measured on its own attempts, and a model call spends energy reading its input tokens as well as generating output tokens. A long tool output lengthens the prompt that the next call reads before generating, so an attempt with long tool outputs and short responses has a higher energy per output token.

\insight[Break-even rule]{Two precisions break even per completed task when they use the same energy to complete it ($\rhoop = 1$). A precision that is cheaper per token stays below break-even only while its per-token saving covers the extra tokens it needs. Formally, the token factor times the boundary factor must stay below $1/\rhotok$. For example, in the business query (\cref{fig:mechanism}b), 4-bit saves 14\% per token, enough to cover up to 16\% more tokens ($1/0.86 = 1.16$), but it needs 59\% more tokens per completed task (all from longer attempts, since both precisions complete all ten), and with a boundary factor of 0.94, the product is $1.59 \times 0.94 = 1.49$, above 1.16, so 4-bit uses 29\% more energy per completed task. Counting only the energy of model calls ($\Einf$) removes the boundary factor, so the token factor alone must stay below $1/\rhotok$ (the dashed curve in \cref{fig:flip}).}

\subsection{Choosing a precision by each measure}
\label{sec:metrics:selection}

Each measure picks a precision for every task. The \emph{per-token choice} is the precision with the lower $\Etok$. The \emph{per-task choice} is the precision with the lower $\Eop$. \cref{tab:rules} covers every pattern of completions, including tasks that one precision or neither completes. A task is \emph{comparable} when both precisions complete it at least three times, so that the choice never rests on one or two lucky attempts. On a comparable task, the two measures \emph{disagree} when the per-token and per-task choices differ. We also use $\Einf$ to locate where a disagreement arises. If $\Einf$ already picks a different precision than $\Etok$, the disagreement comes from the tokens each completion takes (i.e., longer or more failed attempts), and if only $\Eop$ picks a different precision, it comes from tool and waiting energy.

\begin{table}[htb]
\caption{Each pattern of verified completions leads to one per-task choice, and only the first makes a task comparable, so that it enters the counts of disagreements.}
\label{tab:rules}
\begin{tabular}{@{}l l c@{}}
\toprule
\headrow \textbf{Precisions that complete the task} & \textbf{Per-task choice} & \textbf{Comparable} \\
\midrule
Both, at least three times each & Lower $\Eop$ & Yes \\
Only one & That precision & No \\
Both, but one fewer than three times & None & No \\
Neither & None & No \\
\bottomrule
\end{tabular}
\end{table}

On a disagreement, we report the cost of following the per-token choice as its extra energy per completed task (e.g., 30\% means 30\% more than the per-task choice).

\FloatBarrier
\section{Experimental Methodology}
\label{sec:method}
\label{sec:method:overview}

\begin{figure}[t]
\centering
\includegraphics[width=\linewidth]{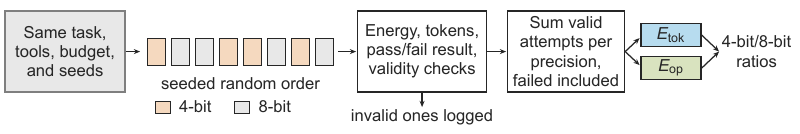}
\caption{In one comparison, every precision runs on each task with the same tools, budget, and seeds, in a seeded random order. Each attempt yields energy, token counts, a pass/fail result, and validity checks. Sums over valid attempts, failed ones included, give each precision's $\Etok$ and $\Eop$.}
\Description{A left-to-right flow. A gray box reading same task, tools, budget, and seeds leads to a row of small orange and gray blocks for 4-bit and 8-bit attempts in seeded random order. An arrow leads to a box reading energy, tokens, pass or fail, and validity checks, with a branch for invalid attempts that are logged. Next, a box reading sum valid attempts per precision, failed included, splits into E tok and E op, which lead to 4-bit over 8-bit ratios.}
\label{fig:design}
\end{figure}

Measuring energy per completed task asks more of an experiment than measuring energy per token, and a fair comparison of two precisions must meet three requirements.
\begin{itemize}
\item \textbf{Isolate precision.} The two precisions of a comparison differ in the model file and share the machine, engine build, tasks, tools, budget, and sampling seeds (\cref{sec:method:models,sec:method:protocol}).
\item \textbf{Verify.} A verifier hardened against known shortcuts decides whether each attempt succeeds, and its own energy stays outside the task window (\cref{sec:method:tasks}).
\item \textbf{Attribute.} Attempts differ in length and outcome, so we attribute energy to each model call and tool call and charge failed attempts to the completions (\cref{sec:method:energy}).
\end{itemize}
\Cref{fig:design} shows how our harness runs one comparison.

\subsection{Testbed and models}
\label{sec:method:models}

\paragraph{\upshape\bfseries Testbed} We run every attempt on bare-metal machines of the Chameleon testbed~\cite{keahey2020chameleon}, so no other tenant shares a machine. Each has two AMD EPYC 7763 processors and one NVIDIA A100 GPU with 40\,GB of memory (80\,GB on one machine). Model and tool work runs on eight cores of the GPU-attached socket, and server memory is bound to that socket's memory (i.e., its NUMA node). These cores use the \texttt{schedutil} frequency governor at 1.5--2.45\,GHz with idle states enabled. Every attempt uses these settings, so host power policy is the same for every precision.

\paragraph{\upshape\bfseries Models} We chose two models that differ in architecture and vendor and that fit entirely in the memory of one 40\,GB GPU at 4-bit and at 8-bit, so all computation stays on the GPU at either precision. The main model, Qwen3-30B-A3B-Instruct-2507~\cite{qwen3modelcard,yang2025qwen3}, is an instruction-tuned mixture-of-experts release without a separate thinking mode. It has 30.5 billion parameters, of which 3.3 billion are active per token (i.e., each token passes through a small subset of expert sub-networks, while the file stores all 30.5 billion). The second model, Devstral Small~2~\cite{devstralmodelcard}, is a dense 24-billion-parameter coding model from another vendor. Both run on llama.cpp~\cite{llamacpp} at one source revision.\footnote{llama.cpp revision \texttt{fcc8915}, built with CUDA on each machine with the same options.} The model files are public GGUF conversions~\cite{bartowski2025qwen3gguf,bartowski2025devstralgguf}, identified by hash and used with the original release's tokenizer and chat template (i.e., the format that turns messages and tool calls into the token sequence the model was trained on).

\paragraph{\upshape\bfseries Precisions} We compare the \Qfour{} and \Qeight{} files, which differ in bit width, quantization scheme, calibration (\cref{sec:background:quant}), and kernels. A developer chooses a file (not a bit width), so we measure the joint effect of all four. The Qwen 16-bit file needs 57\,GiB (\cref{fig:precision}), so we run the 16-bit reference on the 80\,GB machine, where every precision fits entirely in GPU memory at the largest context we use.

\subsection{Tasks and verification}
\label{sec:method:tasks}
\label{sec:method:verify}

Energy per completed task needs a completion criterion that is objective, reproducible, and identical for every precision, so we use \emph{verifiable tasks}, whose result an automatic check accepts or rejects, as repository-level coding benchmarks do~\cite{jimenez2024swebench}. Success is then simply pass or fail, which keeps the variance of a judged or graded score out of the energy comparison. Some problems come in two input sizes, and counting each size as a task gives 29 tasks from 22 problems (\cref{tab:tasks}).

\begin{table}[t]
\caption{An automatic check decides success in every family. Each row shows one example task.}
\label{tab:tasks}
\begin{tabularx}{\linewidth}{@{}>{\raggedright\arraybackslash}p{2.75cm} c >{\raggedright\arraybackslash}X >{\raggedright\arraybackslash}p{3.4cm}@{}}
\toprule
\headrow \textbf{Family} & \textbf{Tasks} & \textbf{Example task} & \textbf{Success check} \\
\midrule
SQL analytics & 6 & Report the number of transactions and the net amount for every customer segment in a sales database, as one structured answer (the business query). & Equals an independent, non-SQL reference \\
\addlinespace[2pt]
Data repair & 2 & Write SQL that keeps only the latest version of each event, drops unpaid events, and totals net amounts by region. & Correct on the given and five further databases \\
\addlinespace[2pt]
Regression control & 2 & Fit a regression per group with the provided tool and report count, slope, intercept, and error. & Matches the reference within $10^{-6}$ \\
\addlinespace[2pt]
Data processing & 4 & Implement a function that removes repeated URLs but keeps records flagged to keep, in input order. & Correct on visible, private, and edge-case inputs \\
\addlinespace[2pt]
Function implementation & 6 & Return the two closest numbers in a list (HumanEval/20), fast enough for large inputs. & Correct on hidden inputs within 2\,s \\
\addlinespace[2pt]
Repository repair & 9 & Find and fix a defect injected into the breadth-first connected-components code of NetworkX. & Matches independent implementations on random graphs, and the package's tests pass \\
\bottomrule
\end{tabularx}
\end{table}

\paragraph{\upshape\bfseries Task families} The tasks cover everyday agent work, from a single SQL query to repository repairs of up to 250 \emph{turns} (i.e., model responses). Each of the six families lets precision change the energy of a task in its own way. We wrote the first three.
\begin{itemize}
\item \textbf{SQL analytics (6).} The agent explores a sales database and submits one structured answer, so one wrong submission fails the attempt.
\item \textbf{Data repair (2).} The agent writes SQL that cleans and totals event records and can test it on a visible database, so failed tests and repairs add energy within the attempt.
\item \textbf{Regression control (2).} The tool does the fitting and the agent reports the result, so energy differences in this family come only from the agent's behavior (its model calls and any repeated tool calls).
\item \textbf{Data processing (4).} The agent implements and tests a pandas or NumPy function. The problems adapt DS-1000~\cite{lai2023ds1000}, a set of data-science problems written against these libraries.
\item \textbf{Function implementation (6).} The agent writes and tests a function that must be correct and finish within a time limit. The problems adapt EvalPerf~\cite{liu2024evalperf}, whose inputs stress running time, with HumanEval+ and MBPP+ inputs~\cite{liu2023evalplus}.
\item \textbf{Repository repair (9).} The agent locates and fixes one injected defect in NetworkX 3.4.2~\cite{hagberg2008networkx}, a widely used Python library for graph algorithms, over many turns of reading, searching, and testing.
\end{itemize}
The adapted tasks use our agent's tools rather than each benchmark's original setup, so their completion rates compare the two precisions under one agent.

\paragraph{\upshape\bfseries Verification} The research verifier alone decides success, so it must be resistant to gaming. It runs once after submission and checks the result against references hidden from the agent, which it computes independently of the agent's approach. It ignores what submitted code reports about its own success. Wherever an agent could pass by hard-coding the answer for the input it was shown, it reruns the check on new inputs generated from private seeds. Before use, every verifier had to accept the reference solutions and reject known shortcuts  (e.g., constant answers, answers hard-coded for the visible input, and forged success messages). During a task, read-only SQL access and a Landlock ruleset\footnote{Landlock is a Linux kernel feature that restricts which files and network connections a process can reach.}~\cite{landlock} keep hidden inputs, expected outputs, and private seeds out of the agent's reach (\cref{app:isolation}).

\subsection{Agent protocol, budgets, and repetitions}
\label{sec:method:protocol}
\label{sec:method:budgets}

\begin{table}[t]
\caption{The expanded budget raises every limit of the primary budget, and decoding is the same under both. Greedy decoding takes the most probable token, which makes the greedy repetition deterministic, and sampled repetitions draw tokens at random with the model-card settings.}
\label{tab:budgets}
\begin{tabularx}{\linewidth}{@{}l r r X@{}}
\toprule
\headrow \textbf{Limit or setting} & \textbf{\shortstack[l]{Primary\\budget}} & \textbf{\shortstack[l]{Expanded\\budget}} & \textbf{Definition} \\
\midrule
\tablegroup{4}{Limits}
Context (tokens) & 16{,}384 & 65{,}536 & Context window of the inference server \\
Turns & 24 & 250 & Model responses, including malformed ones and the submission \\
Tool calls & 48 & 500 & Task-tool calls per attempt \\
Output tokens & 8{,}192 & 65{,}536 & Total over the attempt \\
Output tokens per turn & 2{,}048 & 4{,}096 & Limit of one response \\
Deadline (s) & 600 & 1{,}800 & From attempt start, including model loading and verification \\
\midrule
\tablegroup{4}{Decoding}
Greedy repetition & \multicolumn{2}{c}{temperature 0, top-$k$ 1} & First repetition of each task and precision \\
Sampled, Qwen & \multicolumn{2}{c}{temperature 0.7, top-$p$ 0.8, top-$k$ 20} & Model card~\cite{qwen3modelcard} \\
Sampled, Devstral & \multicolumn{2}{c}{temperature 0.15, top-$p$ 1} & Model card examples~\cite{devstralmodelcard} \\
Seeds & \multicolumn{2}{c}{seed $k$ in repetition $k$} & The same for every precision \\
\bottomrule
\end{tabularx}
\end{table}

\paragraph{\upshape\bfseries Inference server} Each attempt starts a new inference server that serves one request at a time, so no cached state or warm-up carries over between attempts. Within an attempt, the server keeps the computation already done for earlier turns (i.e., the key-value cache) and processes only the new part of each prompt. The harness renders the conversation (including the tool definitions) with the model's own chat template and sends the token identifiers to the native completion endpoint of llama.cpp on the same machine, so it controls exactly which tokens each precision reads.

\paragraph{\upshape\bfseries Turns and tool actions} The system message asks the model to solve one task with the provided tools and to finish by submitting. Each family offers one task tool in place of a general shell, which helps keep hidden material out of the agent's reach. The tool has a few actions (e.g., for SQL analytics, inspecting the schema, running a read-only query, and submitting). Tool calls run one at a time between model calls, each in a new process, so every call's energy is measured separately. The parser accepts only well-formed tool calls in the chat template's format and repairs nothing except a missing closing tag at the end of a response. When a response has no valid call, the harness keeps its text without the rejected call and replies with the parser error, so the response costs a turn at any precision. The harness saves every response, tool call, and tool result and labels each turn (e.g., a malformed call), and \cref{sec:results:rq3} counts agent behaviors from these records.

\paragraph{\upshape\bfseries Setting the budgets} Limits on context, turns, tool calls, output tokens, and time keep every attempt finite, and together they form its \emph{execution budget} (\cref{tab:budgets}). We fixed the primary budget after pilot runs, setting each limit above what pilot attempts outside repository repair needed (e.g., a 600\,s deadline well above the longest pilot attempt, and 2{,}048 tokens per turn for answers of about 1{,}100 tokens). The expanded budget's turn limit equals the step limit of mini-SWE-agent~\cite{minisweagent,yang2024sweagent} for SWE-bench~\cite{jimenez2024swebench}, where a step is also one model response. Its context is the largest at which every precision fits entirely in GPU memory. The model supports contexts of up to 262{,}144 tokens, and the harness keeps the full conversation without truncation or summarization.

Outside repository repair, the limits rarely end a valid sampled attempt. With the primary budget, 389 of 399 attempts on the other families end by submitting, and with the expanded budget, all such attempts end by submitting. However, about two thirds of primary-budget repository attempts end at the context or turn limit for both precisions, which motivates the expanded budget. The limits on tool calls, total output tokens, and time each end at most one valid sampled attempt (\cref{tab:limits} in \cref{app:budgets}).

\paragraph{\upshape\bfseries Repetitions and order} A \emph{repetition} is one attempt on every task at each precision, and repetition $k$ uses seed $k$ for every precision. The first repetition uses greedy decoding, which always takes the most probable next token and makes the attempt repeatable. Later repetitions sample each token at random, using the settings recommended on each model card (\cref{tab:budgets}). Intervals and tests use only these sampled repetitions. To keep slow host changes (such as chip temperature) from favoring either precision, the precisions of each task run in a seeded random order. Attempts run one at a time, with 45\,s between them for power to settle. Any drift in energy per token across repeated identical attempts affects both precisions alike (\cref{app:sensitivity}).

\subsection{Energy measurement and attempt validity}
\label{sec:method:energy}
\label{sec:method:validity}

\paragraph{\upshape\bfseries Energy counters} Every 0.1\,s, we read cumulative energy counters for both CPU packages through Linux powercap~\cite{linuxpowercap} and for the GPU board through the NVIDIA Management Library~\cite{nvml}. The energy of an interval is the difference of two readings. It includes idle energy within the interval and excludes DRAM, storage, fans, and power-supply losses. AMD package counters are estimates from a power model~\cite{schone2021zen}, and the GPU sensor samples power intermittently~\cite{yang2024parttime}, so we compare precisions on the same machine. If a sensor reads high or low by the same factor for both precisions, the 4-bit/8-bit energy ratio stays the same, and so does the precision it favors.

\paragraph{\upshape\bfseries Measured periods} An attempt has three measured periods (\cref{fig:mechanism}a). Model loading and verification are reported in \cref{app:measurement}, and the task window between them gives $O_i$. Within the window, each model call and tool call is a span whose energy is interpolated between the 0.1\,s counter readings, and $I_i$ sums the energy of the model-call spans.

\paragraph{\upshape\bfseries Valid attempts} A \emph{valid} attempt, one whose measurement is complete and undisturbed, passes three checks:
\begin{itemize}
\item Its energy record is complete, and recomputing it from the raw counter readings gives the stored value (\cref{app:measurement}).
\item Its token counts add up (i.e., reused plus newly read input tokens equal the input length).
\item No process it started (e.g., a tool or server process) is still running when it ends, and a monitor that samples CPU use about every 0.6\,s finds no sustained activity by unrelated processes.
\end{itemize}
Invalid attempts are reported in the attempt totals of \cref{tab:testbeds} but are kept out of the energy and completion counts, and none are rerun, since replacing them would let outcomes decide which attempts count. The harness records the cause of every failure (e.g., a wrong answer, a program error or timeout, a missed time limit, or a budget limit). An attempt that the deadline interrupts during generation has an incomplete energy record, because the server stops before the final counter reading, so it is invalid. \Cref{app:sensitivity} counts it as a failure with energy rebuilt from the readings taken before the deadline and shows that the results of \cref{sec:results} still hold. \Cref{tab:testbeds} counts the valid attempts.

\subsection{Conditions}
\label{sec:method:conditions}
\label{sec:method:extensions}

Each row of \cref{tab:testbeds} is a \emph{condition}, one combination of tasks, model, budget, and machine under which we compare the precisions. The \emph{main experiment} runs Qwen at 4-bit and 8-bit on all 29 tasks with the primary budget. The next three conditions test whether its result holds with a larger budget or another model. The \emph{repository rerun} repeats the repository tasks with the expanded budget on the same machine. \emph{Qwen expanded} runs all tasks with the expanded budget and adds the 16-bit reference, which needs the 80\,GB machine. \emph{Devstral} runs all tasks with the second model and the expanded budget. Because a condition can change more than one of model, budget, and machine, we compare precisions within each condition and report differences between conditions descriptively.

\begin{table}[t]
\caption{Each later condition departs from the main experiment (first row) as listed. Repetitions count one greedy repetition plus the sampled ones. Valid gives valid attempts over all attempts.}
\label{tab:testbeds}
\begin{tabularx}{\linewidth}{@{}l l >{\raggedright\arraybackslash}X l c r@{}}
\toprule
\headrow \textbf{Condition} & \textbf{Tasks} & \textbf{Changes from the main experiment} & \textbf{Machine} & \textbf{Repetitions} & \textbf{Valid} \\
\midrule
Main experiment & 29 & None (Qwen, 4-bit and 8-bit, primary budget) & A, 40\,GB & 1+10 & 636/638 \\
Repository rerun & 9 & Repository tasks only, expanded budget & A, 40\,GB & 1+10 & 196/198 \\
Qwen expanded & 29 & Expanded budget, 80\,GB machine, adds 16-bit & D, 80\,GB & 1+6 & 609/609 \\
Devstral & 29 & Devstral with its chat template, sampling, and tool-call syntax, expanded budget & B, 40\,GB & 1+10 & 637/638 \\
Held-out & 77 & BIRD and EvalPerf tasks & C, 40\,GB & 1+2 & 462/462 \\
\bottomrule
\end{tabularx}
\end{table}

\paragraph{\upshape\bfseries Held-out tasks} The held-out tasks extend the comparison to benchmark problems outside our task set. They include 40 simple questions on 11 databases of BIRD~\cite{li2023bird}, a text-to-SQL benchmark over real databases judged by its official execution-match rule~\cite{birdevaluation}. For example, one question asks for the eye color of the superhero whose full name is Karen Beecher-Duncan. They also include the 37 eligible EvalPerf problems~\cite{liu2024evalperf} unused in the main experiment, such as HumanEval/0, which asks whether any two numbers in a list are closer than a threshold. We selected both subsets by rules independent of how any model performed (\cref{app:external}). To test whether the patterns of our task families carry over to new problems, we made two predictions from the main experiment's early results (\cref{sec:method:evolution}) before running any held-out task. Because SQL analytics favored 8-bit and function implementation did not, we predicted that 4-bit would use less energy per completed task on the function-implementation problems of EvalPerf and that 8-bit would match or beat it on the SQL questions of BIRD. \Cref{sec:results:rq4} tests both.

\subsection{Statistical analysis}
\label{sec:method:stats}
\label{sec:method:analysis}

\paragraph{\upshape\bfseries Statistical support} Repeated attempts show whether a choice is more than noise. Each condition uses the sampled repetitions that are complete at every precision, and partly completed repetitions are reported separately (\cref{app:sensitivity}). For each task, we compute the 4-bit/8-bit ratio of each of the three measures over these repetitions. A choice is \emph{statistically supported} when the 95\% bootstrap interval of its ratio lies entirely on the chosen side of one. An interval that includes one supports neither precision. The interval comes from 5{,}000 draws that resample each precision's valid attempts~\cite{efron1993bootstrap}. When only one precision completes a task, the choice rests on completion counts and is supported when an exact test gives $p<0.05$ (McNemar's for attempts paired by seed, Fisher's otherwise~\cite{mcnemar1947note,fisher1935logic}). A disagreement is supported when both of its choices are supported on the same attempts.

\paragraph{\upshape\bfseries Noise baseline} Because every task gets its own test, a few tasks could look like supported disagreements by luck alone, and the noise baseline (i.e., a label-permutation test~\cite{good2005permutation}) measures how many to expect. After the runs, we randomly swap the 4-bit and 8-bit labels among each task's recorded attempts (i.e., in each repetition, a coin flip decides whether the 4-bit and 8-bit attempts trade labels, so the label no longer tells which precision produced it). Each attempt keeps its real energy and outcome, so any disagreement that remains comes from luck alone. Each shuffle yields some number of supported disagreements by luck alone. Repeating the shuffle many times shows how often luck reaches the number we observe, and if it rarely does, the observed count is unlikely to be luck. A second shuffle keeps each task's real per-token gap (\cref{app:stats}).

\paragraph{\upshape\bfseries Summaries across tasks} A condition or family is summarized by its average (i.e., the geometric mean of its task-level ratios), weighting a problem with two input sizes as one problem. Because a geometric mean lets a halving and a doubling cancel, an average near one (i.e., no overall difference) can hide tasks that strongly favor either precision. Intervals for these summaries come from 2{,}000 draws that resample attempts within tasks, or for held-out tasks from 5{,}000 draws that resample tasks and then attempts (\cref{app:stats}).

\paragraph{\upshape\bfseries Confirmatory and exploratory analyses}
\label{sec:method:evolution}
After the pilot runs and before the main experiment, we fixed the tasks, tools, verifiers, primary budget, measured periods, validity rules, and the plan to compare $\Etok$ with $\Eop$. The main experiment tests this plan. After seeing its early results, we added the expanded budget, the held-out tasks with their predictions, and the second model. These extensions are exploratory, and so are later analyses such as the retest on separate repetitions and the model-loading case.

\section{Results}
\label{sec:results}

We answer the four research questions from \cref{sec:intro} in turn, asking whether the two measures favor the same precision (RQ1), how much choosing per token costs (RQ2), where the extra energy comes from (RQ3), and what changes the cheaper precision per completed task (RQ4). All ratios are 4-bit over 8-bit, so values below one favor 4-bit. The three conditions that run all 29 tasks are the main experiment (Qwen with the primary budget), Qwen expanded, and Devstral. A median attempt uses 2--7\,kJ over 15--31\,s, and a held-out attempt uses about 1\,kJ over 12--14\,s (\cref{tab:absolute}).

\subsection{RQ1: \texorpdfstring{Whether the two measures favor the same precision}{Whether the two measures favor the same precision}}
\label{sec:results:rq1}
\label{sec:results:accounting}

\begin{figure}[!t]
\centering
\includegraphics[width=\linewidth]{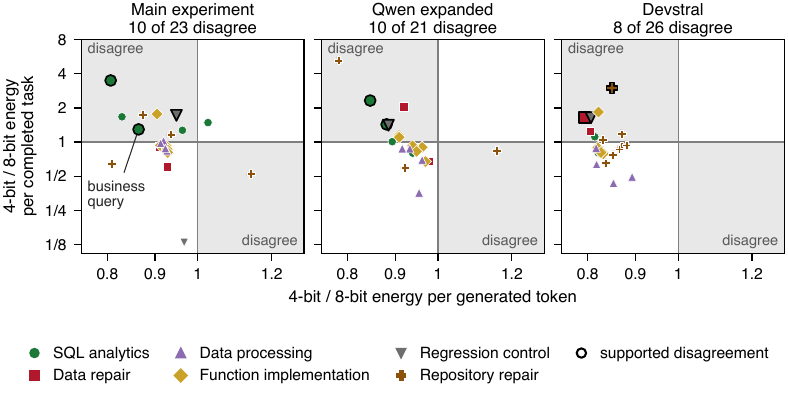}
\caption{4-bit is cheaper per token for nearly every task, and black rings mark the three tasks per condition where it is costlier per completed task beyond run-to-run variation. Each point is a comparable task, placed by its 4-bit/8-bit ratio of energy per output token (horizontal) and of energy per completed task (vertical), on log scales. Points in the two shaded quadrants are disagreements.}
\Description{Three scatter plots, one per condition, with energy per token on the horizontal axis and energy per completed task on the vertical axis, both as 4-bit over 8-bit ratios on log scales. Nearly all points lie left of one on the horizontal axis, while many lie above one on the vertical axis, inside a shaded quadrant labeled disagree.}
\label{fig:choices}
\end{figure}

Energy per token favors 4-bit almost everywhere, yet energy per completed task favors 8-bit on some tasks in all three conditions (\cref{fig:choices}). 4-bit uses 6--16\% less energy per output token than 8-bit on average over the 29 tasks, depending on the condition, and is cheaper on 25 to 29 of them.

Of the 23, 21, and 26 comparable tasks in the three conditions (i.e., tasks that both precisions complete at least three times), the two measures favor different precisions on 10, 10, and 8 tasks respectively. Chance alone could produce most of these, as the noise baseline (\cref{sec:method:analysis}), keeping each task's per-token gap, yields 8.6, 8.2, and 4.8 disagreeing tasks on average (\cref{app:stats}). We thus count only supported disagreements, whose two gaps both exceed run-to-run variation. Three remain in each condition, and at most 3\% of shuffles produce three or more, so chance alone rarely yields this many. For Qwen, these are the business query at both input sizes (8{,}192 and one million rows) and one regression-control task. For Devstral, these are one data-repair, one regression-control, and one repository-repair task. The same three Qwen tasks recur at both budgets, on different machines and separate repetitions, so they reflect the tasks rather than one machine or one set of runs.

The other 6, 8, and 3 tasks (29 minus the 23, 21, and 26 comparable ones) are the hardest, as at least one precision completes them fewer than three times, so we compare completion counts instead. For Qwen, the budget decides which one completes more (\cref{sec:results:rq4}). Under the primary budget, 8-bit completes the hardest repairs more often (7 completions against 2 for 4-bit in the main experiment), and under the expanded budget, 4-bit does (13 against 3 in Qwen expanded).

\paragraph{\upshape\bfseries Model loading} Even in the worst case of \cref{sec:metrics:defs}, one model load per attempt, the business query at 8{,}192 rows stays a supported disagreement for Qwen at both budgets, and for Devstral, the regression-control task at 8{,}192 rows and the repository-repair task also stay supported disagreements. A 4-bit load takes 0.4--0.7\,kJ less than an 8-bit one (\cref{app:measurement}). Loading thus moves the condition averages (i.e., geometric means of the task-level ratios) per completed task toward 4-bit by 0.02 to 0.12, and removes some disagreements without creating any.

\insight[Finding 1]{A precision that is cheaper per token can cost more per completed task. 4-bit uses 6--16\% less energy per token on average, yet in each condition, three tasks cost more per completed task at 4-bit by more than run-to-run variation, such as the SQL business query for Qwen and the repository repair for Devstral (\cref{fig:choices}). On the business query at one million rows, for example, 4-bit uses 14\% less energy per token but 29\% more per completed task (\cref{fig:mechanism}b).}

\begin{table}[t]
\caption{4-bit has the lower median energy per token in the three conditions that run all 29 tasks, while medians per attempt and per completed task differ little, and wide interquartile ranges show long attempts in the expanded-budget conditions. Entries are medians over valid sampled attempts, with interquartile ranges in parentheses. Energy per completed task is the median across tasks that both precisions complete at least once. Medians pool tasks, so they can differ from the task-level ratios of \cref{fig:settings}.}
\label{tab:absolute}
\setlength{\tabcolsep}{3.5pt}
\begin{tabular}{@{}lrrrrr@{}}
\toprule
\headrow \textbf{Condition, precision} & \textbf{J per token} & \textbf{\shortstack[r]{kJ per\\completed task}} & \textbf{\shortstack[r]{kJ per\\attempt}} & \textbf{\shortstack[r]{s per\\attempt}} & \textbf{\shortstack[r]{Output tokens\\per attempt}} \\
\midrule
Main experiment, 4-bit & 2.38 {\scriptsize(2.24--2.98)} & 2.5 {\scriptsize(1.3--5.3)} & 2.0 {\scriptsize(1.2--3.8)} & 16 {\scriptsize(13--22)} & 501 {\scriptsize(391--823)} \\
Main experiment, 8-bit & 2.64 {\scriptsize(2.46--3.20)} & 2.2 {\scriptsize(1.5--8.3)} & 2.3 {\scriptsize(1.3--5.2)} & 19 {\scriptsize(16--29)} & 501 {\scriptsize(374--836)} \\
Qwen expanded, 4-bit & 2.69 {\scriptsize(2.50--3.53)} & 2.0 {\scriptsize(1.4--3.7)} & 2.1 {\scriptsize(1.3--16.5)} & 15 {\scriptsize(13--125)} & 574 {\scriptsize(417--869)} \\
Qwen expanded, 8-bit & 2.92 {\scriptsize(2.71--3.88)} & 2.0 {\scriptsize(1.4--3.5)} & 2.0 {\scriptsize(1.4--17.1)} & 18 {\scriptsize(16--129)} & 558 {\scriptsize(397--1195)} \\
Devstral, 4-bit & 10.1 {\scriptsize(9.9--11.5)} & 7.5 {\scriptsize(3.9--35.3)} & 6.6 {\scriptsize(3.3--28.3)} & 30 {\scriptsize(19--173)} & 576 {\scriptsize(312--1336)} \\
Devstral, 8-bit & 12.2 {\scriptsize(12.0--13.7)} & 7.7 {\scriptsize(4.4--34.7)} & 6.7 {\scriptsize(3.9--28.6)} & 31 {\scriptsize(23--173)} & 520 {\scriptsize(297--1307)} \\
Held-out, 4-bit & 3.01 {\scriptsize(2.39--3.59)} & 0.86 {\scriptsize(0.77--1.22)} & 0.90 {\scriptsize(0.76--1.19)} & 12 {\scriptsize(11--13)} & 238 {\scriptsize(160--391)} \\
Held-out, 8-bit & 2.96 {\scriptsize(2.60--3.77)} & 0.96 {\scriptsize(0.81--1.34)} & 0.99 {\scriptsize(0.78--1.31)} & 14 {\scriptsize(13--15)} & 262 {\scriptsize(165--396)} \\
\bottomrule
\end{tabular}
\end{table}

\subsection{RQ2: The cost of choosing by energy per token}
\label{sec:results:rq2}

Where the two measures disagree, following energy per token is expensive. On the supported disagreements, the per-token choice (i.e., the precision with the lower energy per token, \cref{sec:metrics:selection}) uses a median of 66\% more energy per completed task than the per-task choice (i.e., the precision with the lower energy per completed task, \cref{fig:choices}). It ranges from 29\% more on Qwen's business query at one million rows to 3.5 times as much at 8{,}192 rows. 8-bit spends about the same energy per completed task at both sizes, so the larger cost at 8{,}192 rows comes from 4-bit, which took eight turns per attempt rather than three and generated about four times as many tokens as 8-bit.

Condition averages (\cref{sec:method:analysis}) dilute this cost, because tasks that favor each precision offset each other. The average 4-bit/8-bit ratio per completed task is 1.17 in the main experiment (17\% more for 4-bit) and 0.92 in Qwen expanded and for Devstral (8\% less). Such an average suggests a modest difference either way, while single tasks cost up to 3.5 times as much.

\paragraph{\upshape\bfseries Retest on separate repetitions} To check that the cost holds on new runs, as it would for an operator who chooses a precision from past runs, we find supported disagreements using one half of the repetitions and measure their cost on the other half, then swap the halves. For Qwen, the per-token choice still costs a median of 31\% and 45\% more per completed task at the two budgets, less than the 66\% above but still large. The per-task choice stays cheaper in 10 of 11 cases. For Devstral, the per-task choice stays cheaper in 2 of 5.

\insight[Finding 2]{Choosing by energy per token can be costly. On the tasks of Finding 1, the precision that is cheaper per token uses a median of 66\% more energy per completed task, and up to 3.5 times as much. For Qwen, the extra cost persists at a median of 31--45\% on separate repetitions, and averages over all tasks dilute it (\cref{fig:choices}).}

\subsection{RQ3: Where the extra energy comes from}
\label{sec:results:rq3}

\begin{figure}[t]
\centering
\includegraphics[width=0.75\linewidth]{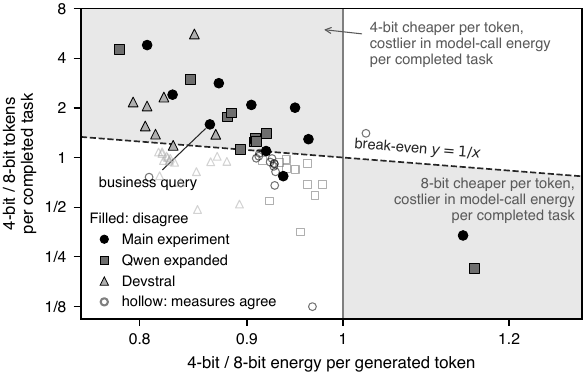}
\caption{Extra tokens outweigh the per-token saving. Filled markers, the disagreeing tasks of \cref{fig:choices}, lie almost all in the shaded regions. Each comparable task is placed by its 4-bit/8-bit ratio of energy per output token (horizontal) and of tokens per completed task (vertical), on log scales. On the dashed curve $y=1/x$, the two precisions break even in model-call energy per completed task. Shaded regions mark where that energy and energy per token favor different precisions.}
\Description{A scatter plot on log scales. The horizontal axis is the 4-bit over 8-bit energy per token, mostly between 0.8 and 1. The vertical axis is the 4-bit over 8-bit tokens per completed task, from about one eighth to eight. A dashed curve y equals one over x marks break-even for model-call energy per completed task. The region left of one and above the curve and the region right of one and below it are shaded. Filled markers in three shades of gray, the disagreeing tasks of three conditions, lie almost all in the shaded regions, while hollow markers for agreeing tasks lie near the curve. The business query is labeled.}
\label{fig:flip}
\end{figure}

The extra energy comes from extra tokens (\cref{fig:flip}). Across all disagreeing tasks, supported or not (\cref{eq:product} holds exactly on each), the precision that is cheaper per token needs a median of 1.6 to 2.0 times as many tokens per completed task in the three conditions (the token factor of \cref{eq:product}). A per-token saving of 6--16\% covers at most 6--19\% extra tokens (\cref{sec:metrics:identity}), so these token factors lie far past break-even. Per-token ratios cluster between 0.8 and 1, while the token factor varies by more than ten times, so tokens per completed task largely decide which precision is cheaper. Tool and waiting energy (the boundary factor) changes the per-task choice on at most one disagreeing task per condition, so nearly all disagreements arise inside the model calls.

\paragraph{\upshape\bfseries Longer attempts and failed ones} Under the primary budget, failed attempts (the second term of \cref{eq:failures}) carry the bulk of the extra energy per completed task in most disagreeing tasks (7 of 10), and under the expanded budget, longer attempts carry that extra energy. On SQL analytics in the main experiment, Qwen submits a wrong answer about three times as often at 4-bit as it does at 8-bit (20 rather than 6 out of 60 attempts). Even its successful 4-bit attempts generate about 1.8 times as many tokens. In the 18 disagreeing tasks of Qwen expanded and Devstral, the precision that is cheaper per token generates a median of 1.5 to 1.8 times as many tokens per attempt (the length factor of \cref{eq:tokens}), but completes as often (a median completion factor of one). In these tasks, the extra energy comes mainly from longer attempts.

\paragraph{\upshape\bfseries What the agents did differently} In the saved traces, the extra tokens include malformed output, narrated plans (i.e., text describing the next step before each tool call), and repeated tests. For SQL analytics, Qwen's tool call is missing or malformed nine times as often at 4-bit as it does at 8-bit (27 rather than 3 out of 60 attempts). Even when both precisions always succeed, 4-bit can work less efficiently. In the business query of \cref{fig:mechanism}b, all 20 attempts ran the same schema lookup and the same single query and ended with the same correct answer. However, the 4-bit model used about 60\% more output tokens (430 on average, where 8-bit used 271 every time). In most of its attempts (6 of 10), it first submitted a numerically correct answer one closing brace short, which the parser rejected, so it submitted again, and in 3 others it described its plan before each call. Devstral at 4-bit drops the same brace just as often on a regression-control task (6 of 10 attempts), which helps explain why 8-bit is cheaper on regression control. In a repository repair that Devstral completed in all ten attempts at both precisions, 4-bit took about 70\% more turns (a median of 18.5 rather than 11) and re-ran already passing tests about twice as often (3.3 rather than 1.7 times per attempt).

\insight[Finding 3]{Extra tokens drive the cost. Where the two measures disagree, the precision that is cheaper per token needs a median of 1.6 to 2.0 times as many tokens per completed task, more than its per-token saving can cover (\cref{fig:flip}). The extra tokens come from (1) longer attempts, as malformed tool calls, narrated plans, and repeated tests add tokens even to successful attempts, and (2) failed attempts, which dominate the extra energy on most of these tasks.}

\subsection{RQ4: \texorpdfstring{What changes the cheaper precision per completed task}{What changes the cheaper precision per completed task}}
\label{sec:results:rq4}

\begin{figure}[t]
\centering
\includegraphics[width=0.75\linewidth]{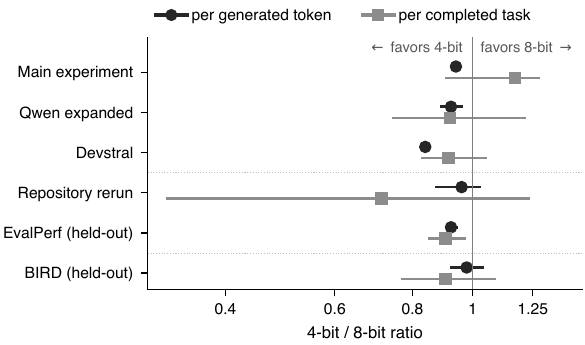}
\caption{The per-token ratio (circles) lies below one in every condition, while the per-task ratio (squares) falls on both sides of one. Each row shows geometric-mean 4-bit/8-bit ratios over problems, and the per-task ratio is taken over the problems that both precisions complete. Bars are 95\% bootstrap intervals from 2{,}000 draws over attempts (5{,}000 draws over tasks and attempts for EvalPerf and BIRD).}
\Description{A dot plot with one row per condition. Dark circles for energy per token lie left of one in every row, with narrow intervals except for the repository rerun and BIRD, whose intervals reach one. Gray squares for energy per completed task lie on both sides of one with wide intervals that cross one, except for EvalPerf. BIRD and EvalPerf appear as separate rows.}
\label{fig:settings}
\end{figure}

Across conditions, 4-bit's per-token saving persists while the cheaper precision per completed task varies (\cref{fig:settings}). The per-token ratio lies below one in every condition (clearly so in all except BIRD and the repository rerun, which repeats the nine repository repairs with the expanded budget). However, per completed task, only the EvalPerf condition average clearly favors 4-bit. The tasks where 4-bit costs more per completed task differ between the two models, and for Qwen, the budget decides which precision completes more of the hardest repairs.

\begin{table}[t]
\caption{The family that favors each precision changes across conditions. Shaded bold cells mark families whose 95\% bootstrap interval (2{,}000 draws) excludes one, orange where 4-bit is favored and blue where 8-bit is favored. Entries are family-level 4-bit/8-bit ratios of energy per completed task, as geometric means over the problems of each family.}
\label{tab:families}
\begin{tabular}{@{}lccc@{}}
\toprule
\headrow \textbf{Family} & \textbf{Main experiment} & \textbf{Qwen expanded} & \textbf{Devstral} \\
\midrule
SQL analytics & \cellcolor{favoreight}\textbf{1.74} & 1.13 & \cellcolor{favorfour}\textbf{0.88} \\
Data repair & 0.74 & 1.18 & \cellcolor{favoreight}\textbf{1.44} \\
Regression control & 0.47 & \cellcolor{favoreight}\textbf{1.25} & \cellcolor{favoreight}\textbf{1.18} \\
Data processing & 0.93 & \cellcolor{favorfour}\textbf{0.66} & \cellcolor{favorfour}\textbf{0.59} \\
Function implementation & 0.99 & \cellcolor{favorfour}\textbf{0.92} & 0.96 \\
Repository repair & 1.54 & 0.83 & 0.91 \\
\bottomrule
\end{tabular}
\end{table}

\paragraph{\upshape\bfseries Task families} Families favor different precisions across conditions (\cref{tab:families}). SQL analytics favors 8-bit in the main experiment but favors 4-bit for Devstral, regression control favors 8-bit in Qwen expanded and for Devstral, and data processing favors 4-bit in Qwen expanded and for Devstral. In the main experiment, the regression-control average (0.47) hides a supported disagreement, because 4-bit uses far less energy per completed task at 8{,}192 rows (0.13, about an eighth of 8-bit's) but uses more at one million rows (1.72, or 72\% more).

\paragraph{\upshape\bfseries Expanded budget} With the expanded budget, Qwen at 4-bit completes more repository repairs, while per completed task the two precisions stay within run-to-run variation. Summed over the nine repairs, both precisions complete 28 under the primary budget (out of 89 and 90 attempts), with 8-bit ahead on the hardest ones (\cref{sec:results:rq1}). About two thirds of these attempts end at the turn or context limit (\cref{sec:method:budgets}). However, in the repository rerun, 4-bit completes about 70\% more repairs than 8-bit (45 rather than 26 out of 89, $p=0.006$, \cref{app:sensitivity}), and Qwen expanded shows the same gain (27 rather than 16 out of 54). Nearly half of the rerun's gain comes from one repair that 4-bit completed in 9 of 10 attempts and 8-bit completed in none, because 8-bit kept working until it reached the turn or context limit (about 220 turns per attempt on average, against about 30 for 4-bit). Devstral, run only with the expanded budget, completes 75 out of 90 repair attempts at each precision. Per completed task, the rerun's ratio is 0.71 [0.32, 1.24], or about 29\% less energy for 4-bit.

\paragraph{\upshape\bfseries 16-bit reference} In Qwen expanded, the condition that adds the 16-bit reference, 16-bit uses 31\% more energy per token than 8-bit and 14\% more per completed task [0.93, 1.49]. Quantizing from 16-bit to 8-bit thus saves energy per token, and per completed task it leaves a 14\% gap within run-to-run variation. Energy per token picks a quantized file on every task, and the three tasks where the per-task choice picks 16-bit are all within run-to-run variation.

\paragraph{\upshape\bfseries Held-out tasks} For Qwen, the patterns of our task families carry over to new benchmark problems, as both predictions stated in \cref{sec:method:extensions} hold. On EvalPerf, 4-bit uses less energy per completed task (0.90 [0.85, 0.98], or 10\% less) and completes about as often (65 out of 74 attempts at 4-bit and 68 out of 74 at 8-bit). A partly finished repetition on another machine widens the interval to [0.79, 1.01] (\cref{app:sensitivity}). On BIRD, as predicted, neither precision is supported per completed task (0.91 [0.77, 1.09]).

\insight[Finding 4]{Across a second model, a larger budget, and held-out tasks, 4-bit stays cheaper per token, but the task families that favor each precision per completed task change. SQL analytics favors 8-bit in the main experiment (1.74) but 4-bit for Devstral (0.88), data processing favors 4-bit under the expanded budget (0.66 and 0.59), and held-out benchmarks match the predictions made before they ran (EvalPerf 0.90). 16-bit uses 31\% more energy per token than 8-bit and is never clearly cheaper per completed task (\cref{fig:settings}, \cref{tab:families}).}

\FloatBarrier
\section{Discussion}
\label{sec:discussion}

\subsection{Choosing a precision in practice}
\label{sec:discussion:implications}

A deployment can check which precision is cheaper per completed task from token counts and the per-token saving. By the break-even rule (\cref{sec:metrics:identity}), 4-bit stays cheaper only while its per-token saving covers the extra tokens it needs, and a saving of 6--16\% covers at most 6--19\% more tokens per completed task in model-call energy. We thus suggest that developers track tokens per completed task to see which precision is cheaper, using the logs of repeated attempts at each precision.

\paragraph{\upshape\bfseries For developers of local agents}
\begin{itemize}
\item \textbf{Short coding functions.} A developer whose agent mostly writes and tests short functions can start with 4-bit. On our function-implementation and held-out EvalPerf tasks, 4-bit used the same or less energy per completed task on average.
\item \textbf{Structured SQL answers.} A developer whose analytics agent answers business questions with one structured SQL result should compare both precisions on a sample of their own questions, with several attempts each. In our SQL analytics family, 8-bit used less energy per completed task in one condition and 4-bit in another (\cref{tab:families}).
\item \textbf{Long repository sessions.} In long sessions with a large context window and many turns, 4-bit completed at least as many repairs as 8-bit in every condition with the expanded budget (\cref{sec:results:rq4}). A developer whose agent works through long sessions on a code base can thus start with 4-bit and switch to 8-bit where 4-bit's extra tokens per completed task exceed what its per-token saving covers on their own tasks.
\end{itemize}

\paragraph{\upshape\bfseries For benchmark and measurement designers} Five practices make agent energy results comparable across studies, and per-attempt records let others recompute both measures under another boundary.
\begin{enumerate}
\item Report energy per completed task beside energy per output token.
\item Charge failed attempts to the completions.
\item Repeat attempts, and compare per-task differences with run-to-run variation.
\item State the energy boundary (e.g., CPU packages and GPU board, or wall power), whether model loading is included, the budget, and the number of repetitions.
\item Include the tests the agent runs while working in its task energy, and exclude the separate check that scores its final result.
\end{enumerate}

\subsection{Scope and limitations}
\label{sec:discussion:limits}

\paragraph{\upshape\bfseries Deployment setting} We study one agent that sends one request at a time to an open-weight model of 24 or 30 billion parameters, resident on a dedicated GPU and served by one engine, with model loading excluded as for a server that stays loaded (\cref{sec:metrics:defs}). This setting charges all energy in the task window to one task and keeps model calls, tools, and host activity separable. Shared or batched serving would divide idle and host energy among tenants. Our machines cover two memory sizes of one GPU generation and one processor family.

\paragraph{\upshape\bfseries Agent interface and budgets} Each task family offers a few tool actions in place of a general shell, and a response with a malformed tool call costs a turn (\cref{sec:method:protocol}). This strict protocol charges each precision for its own malformed output under the same rule, which is one source of the extra tokens (\cref{sec:results:rq3}). A deployment could reduce this cost for both precisions by repairing malformed calls or by constraining the output format with a llama.cpp grammar, and energy per completed task would show how much it saves. Both precisions share one budget, as under a single deployment policy, and a budget set for each precision could give 4-bit the longer context that its smaller file allows.

\paragraph{\upshape\bfseries Task set and verification} We study a fixed set of verifiable tasks, because an automatic check, as in SWE-bench and BIRD~\cite{jimenez2024swebench,li2023bird}, makes completion directly measurable at every precision. The held-out benchmarks test whether the patterns of our task families carry over beyond this set (\cref{sec:results:rq4}). A repository patch counts only when it passes the verifier, as in SWE-bench, so the energy of partly correct patches is charged to the completions. Open-ended work such as web interaction, research, or planning would add judged or graded outcomes. Our tasks represent the short and medium-length work of a local agent, and the same per-task accounting applies to multi-hour projects and to models with a thinking mode, which raises energy per problem at nearly unchanged energy per token~\cite{luo2026blackwell}. On longer and harder tasks, where completion rates fall, failed attempts weigh more in each completion's energy (\cref{eq:failures}), so the per-task measure matters more there. The precisions of one model share their base weights, so both saw the same training data, and privately seeded verifier inputs limit what a model gains from having seen public benchmark problems during training.

\subsection{Reproducibility}
\label{sec:reproducibility}
\label{sec:discussion:reproducibility}

The per-attempt records let others recompute every reported energy and completion figure. Each attempt with a complete measurement record stores its settings, raw energy-counter readings, the time span of each model and tool call, token counts, the verifier decision, and its validity checks, and every incomplete or invalid attempt is logged with its reason. File hashes identify the plans, model files, task assets, and the analysis scripts behind the reported intervals and figures, and one manifest file records the machine, tool interface, and harness version of each condition. Upon acceptance, we will release the code, task definitions and verifiers, run settings, and per-attempt records on Zenodo, and identify by source and version any third-party asset whose license prevents redistribution.

\FloatBarrier
\section{Conclusion}
\label{sec:conclusion}

Quantization is making it practical to run agents on a single local GPU, and we asked whether a precision that is cheaper per token is also cheaper per completed task, a question for quantized models in general. Measuring both on the same attempts of locally hosted coding and data agents, we find that 4-bit is cheaper per token almost everywhere, while the cheaper precision per completed task depends on the task, the model, and the execution budget. We trace the difference to extra tokens from malformed tool calls, narrated plans, repeated tests, and failed attempts, which can outweigh the per-token saving. We suggest that developers track tokens per completed task over repeated attempts at each precision, and that benchmark designers report energy per completed task beside energy per token. For anyone who hosts an agent, the energy bill arrives per completed task, not per token.

\bibliographystyle{ACM-Reference-Format}
\bibliography{\sourcepath/sections/refs}

\clearpage
\appendix
\section*{Appendix}
\section{Verifier isolation and adversarial checks}
\label[appendix]{app:isolation}

Expected answers, reference queries, and private seeds stay outside the agent workspace (\cref{sec:method:verify}), and the verifier draws further inputs from the same private seeds for every precision. The repository verifier compares library outputs with independent implementations on 32 random graphs. Submitted functions and repository-verifier code run in sandboxed child processes (Landlock~\cite{landlock}) that deny file creation, process creation, and TCP connections. Data-processing functions run under a syntactic allow-list, and SQL tools deny writes and access to catalogs, pragmas, and attached databases. The checks of \cref{tab:adversarial} test each kind of interface the agent is given.

\begin{table}[!htb]
\caption{Every adversarial check had to pass before the experiment that uses the verifier.}
\label{tab:adversarial}
\begin{tabularx}{\linewidth}{@{}>{\raggedright\arraybackslash}p{3.1cm} >{\raggedright\arraybackslash}X >{\raggedright\arraybackslash}p{2.9cm}@{}}
\toprule
\headrow \textbf{Failure mode tested} & \textbf{Attempted by the check} & \textbf{Required outcome} \\
\midrule
Reading hidden material & Answer-file reads, directory traversal, parent-environment reads, symlinks & Denied, planted marker unseen \\
Escaping the sandbox & File writes, process creation, TCP connections, disallowed imports & Denied, planted marker unseen \\
Claiming success & Forged success messages, premature successful exits & Rejected by the verifier \\
Answering without solving & Constant answers, hard-coded visible answers, lookups of previously exposed hidden inputs & Rejected by the verifier \\
Weak repository verifier & Original library and the library with its injected defect & Accepts the original, rejects the defect \\
Weak reference & Reference solutions, wrong queries, unmodified stubs & Accepts references, rejects the rest \\
\bottomrule
\end{tabularx}
\end{table}

\section{How attempts ended under each budget}
\label[appendix]{app:budgets}

Limits end attempts mainly in repository repair (\cref{sec:method:budgets}, \cref{tab:limits}). Each model call may generate at most the smaller of the per-turn limit and the unused part of the total, and an attempt ends with a \emph{context stop}, without sending a request, when its prompt plus this allowance would exceed the context. The deadline counts from the start of the attempt, so it includes 8 to 12\,s of model loading with the primary budget and 8 to 17\,s with the expanded budget, at most 2\% of either deadline. Every valid sampled attempt in the main experiment ended before the 600\,s deadline, and the expanded deadline of 1{,}800\,s ended one valid sampled attempt, a 4-bit repository attempt in Qwen expanded after 97 turns.

\begin{table}[!htb]
\caption{Outside repository repair, nearly all sampled attempts submit. Entries are percent of valid attempts, 4-bit / 8-bit. With the expanded budget, all attempts on the other families submitted in Qwen expanded and Devstral. All attempts stayed within the tool-call limit, and one reached the total output-token limit.}
\label{tab:limits}
\begin{tabular}{@{}l l c c c c@{}}
\toprule
\headrow \textbf{Condition} & \textbf{Tasks} & \textbf{Submitted} & \textbf{Context limit} & \textbf{Turn limit} & \textbf{Per-turn token limit} \\
\midrule
Main experiment & Other families & 97 / 98 & 0 / 0 & 2 / 2 & 0 / 1 \\
 & Repository & 34 / 34 & 40 / 38 & 25 / 28 & 1 / 0 \\
Qwen expanded & Repository & 52 / 35 & 22 / 31 & 19 / 33 & 6 / 0 \\
Repository rerun & Repository & 54 / 33 & 27 / 33 & 16 / 31 & 3 / 3 \\
Devstral & Repository & 87 / 86 & 6 / 6 & 1 / 0 & 7 / 9 \\
Held-out & BIRD & 92 / 92 & 1 / 0 & 6 / 8 & 0 / 0 \\
 & EvalPerf & 100 / 99 & 0 / 0 & 0 / 0 & 0 / 1 \\
\bottomrule
\end{tabular}
\end{table}

\section{Energy-record checks, model loading, and verification energy}
\label[appendix]{app:measurement}

Every stored energy value is recomputed from its raw counter readings (\cref{sec:method:validity}). A decrease of a CPU package counter~\cite{linuxpowercap} counts as a wrap (i.e., an overflow back to zero) only when it falls from the top quarter of its range of about 65.5\,kJ to the bottom quarter, and any other decrease invalidates the attempt. The recomputed energy of each period must match the stored value to a relative tolerance of $10^{-12}$.

A 4-bit load takes 0.4 to 0.7\,kJ less than an 8-bit one, with median loading energy per sampled attempt of 1.55 to 2.14\,kJ for 4-bit and 2.00 to 2.87\,kJ for 8-bit across the conditions. Because loading costs about as much as a short attempt (0.07 to 1.88 times the median operational energy of an attempt for 4-bit and 0.05 to 2.40 times for 8-bit), \cref{sec:results:rq1} charges one load to every attempt as a worst case, and supported disagreements remain for both models. The research verifier stays outside the agent's energy (\cref{sec:method:energy}), and its median energy is at most 0.13\,kJ, rising to 10.0\,kJ for 4-bit attempts in the repository rerun, where submitted patches run the test suite.

\section{Sensitivity analyses and energy drift}
\label[appendix]{app:sensitivity}
\label[appendix]{app:validity}

\Cref{tab:sensitivity} confirms the results of \cref{sec:results} under other handling of remote logins, missing energy records, partial repetitions, pairing, and answer matching.

\paragraph{\upshape\bfseries Missing energy records} All three incomplete energy records support the same conclusions once rebuilt (\cref{sec:method:validity}). \Cref{tab:sensitivity} rebuilds the two that the deadline cut off during generation from the readings taken before it, and bounds the 8-bit record lost after a verified completion by the task's range.

\paragraph{\upshape\bfseries Partial repetitions} Two conditions have a partial repetition (\cref{sec:method:analysis}), the eighth in Qwen expanded and a held-out repetition on machine A. Held-out repetitions were assigned to machines before any ran, and the analysis uses machine C, where both sampled repetitions finished.

\begin{table}[!htb]
\caption{The reported conclusions hold under each alternative analysis. Ratios are 4-bit/8-bit energy per completed task with 95\% bootstrap intervals, and disagreement counts use comparable tasks.}
\label{tab:sensitivity}
\begin{tabularx}{\linewidth}{@{}>{\raggedright\arraybackslash}p{3.9cm} >{\raggedright\arraybackslash}X >{\raggedright\arraybackslash}p{2.0cm}@{}}
\toprule
\headrow \textbf{Condition and main analysis} & \textbf{Alternative result} & \textbf{Conclusion} \\
\midrule
Repository rerun, attempts overlapping brief remote logins kept & Without them 45 of 85 against 23 of 83 completions ($p=0.001$), 0.61 [0.27, 1.11] & Holds \\
Repository rerun, 4-bit deadline failure without energy & Rebuilt energy moves one task to 8-bit, 3 of 4 disagreements, against 2 of 4 in the main analysis, 0.88 [0.39, 1.63] & Holds \\
Devstral, 8-bit deadline failure without energy & Rebuilt energy 0.86 [0.79, 1.02], 8 of 26 disagreements & Holds \\
Main experiment, 8-bit record lost after a verified completion & Energy bounded by the task's range gives 1.165 to 1.173, 10 of 23 disagreements & Holds \\
Qwen expanded, eighth repetition, run on another machine, excluded & With it, per token 0.93 [0.89, 0.96], per task 1.03 [0.78, 1.29], 12 of 23 disagreements, 4 supported, 16-bit/8-bit per task 1.13 [0.88, 1.36] & Holds, 4 supported \\
EvalPerf (C), partial repetition on A excluded & With it, 0.88 [0.79, 1.01], 82 of 92 against 84 of 91 completions & Same direction, as in \cref{sec:results:rq4} \\
BIRD (C), partial repetition on A excluded & With it, 0.96 [0.79, 1.15] & Holds \\
Held-out (A), partial repetition alone (17 EvalPerf, 40 BIRD tasks) & EvalPerf 0.85 [0.71, 0.98], BIRD 0.95 [0.84, 1.05] & Holds \\
Three 29-task conditions, unpaired attempts & Only repetitions valid at both precisions, 1.16 [0.89, 1.28], 0.92 [0.74, 1.21], 0.92 [0.83, 1.06], 10 of 23, 10 of 21, 8 of 26 disagreements & Holds \\
BIRD (C), answers compared as sets~\cite{birdevaluation} & As multisets or ordered lists, both 44 of 80 against 52 of 80 completions, 0.96 [0.82, 1.15] & Holds \\
\bottomrule
\end{tabularx}
\end{table}

\paragraph{\upshape\bfseries Drift} Any drift in energy per token across repetitions affects both precisions alike (\cref{sec:method:protocol}). For attempts of the same task and precision with identical per-call token counts in every sampled repetition (105, 68, and 182 attempts on machines A, D, and B), energy per output token changed by $-0.06$\%, $+0.35$\%, and $-0.02$\% per repetition. The 4-bit and 8-bit slopes differed by less than 0.1\% per repetition on every machine, with 95\% intervals within 0.2\% on machines A and B and within 0.75\% on machine D. Machine D swings by about 3\% between repetitions, but both precisions run in the same repetition, so the swings cancel in the ratio.

\section{Held-out task selection and statistical procedures}
\label[appendix]{app:external}
\label[appendix]{app:stats}

Both held-out subsets were fixed by rules independent of how any model performed (\cref{sec:method:extensions}).

\paragraph{\upshape\bfseries Function-implementation problems} Of the 112 EvalPerf~\cite{liu2024evalperf} problems outside the main set, 37 were eligible and all 37 were used (58 lacked the runtime separation, 15 had an incorrect reference solution, and two lacked computable canonical outputs). A problem is eligible if its canonical, fastest, and slowest reference solutions are correct and the slowest takes at least three times the time limit (the greater of 2\,s and three times the canonical solution's median runtime), so that a correct but slow solution fails. Every failure had a cause other than the time limit or the $10^{-6}$ regression tolerance, so both thresholds leave the reported comparisons unchanged.

\paragraph{\upshape\bfseries BIRD questions} A seeded shuffle (seed 20260929) of the 925 BIRD~\cite{li2023bird} questions labeled simple in the June 2024 development release chose the first 40 with at most five per database, giving 11 databases.

\paragraph{\upshape\bfseries Intervals} Condition and family summaries (\cref{sec:method:analysis}) resample attempts within tasks (2{,}000 draws~\cite{efron1993bootstrap}), and held-out summaries resample tasks and then attempts (5{,}000 draws), grouped by database for BIRD. Unpaired draws resample each precision's valid attempts, paired draws resample whole repetitions so both precisions stay together, and draws without a completion are omitted.

\paragraph{\upshape\bfseries Noise baseline} The baseline (\cref{sec:method:analysis}) counts the disagreements expected by luck alone~\cite{good2005permutation}. The full shuffle swaps the two precision labels with probability 1/2 in each repetition valid at both precisions. The second shuffle divides each 4-bit attempt's model-call energy by the task's per-token ratio, swaps the labels, and multiplies the attempts now labeled 4-bit by that ratio, so only the per-token gap depends on precision. Supported counts use 300 shuffles with 1{,}000 bootstrap draws each, and disagreement counts use 10{,}000 shuffles. The second shuffle yields 8.6, 8.2, and 4.8 disagreements on average, against the 10, 10, and 8 disagreements observed. Both shuffles yield fewer than one supported disagreement on average, and three or more appear in at most 2.7\% of full shuffles and 1.7\% of second shuffles.

\end{document}